\documentclass[a4paper,reqno]{amsart}
\usepackage[text={5.0in,8in},centering]{geometry}
\usepackage{amsrefs}
\usepackage{amssymb}
\usepackage{mathrsfs}

\usepackage{xspace}
\usepackage{mathtools}

\usepackage{enumitem}

\usepackage{color}
\definecolor{trustcolor}{rgb}{0.71,0.14,0.07}

\usepackage{hyperref}

\usepackage{pgf}
\usepackage{tikz}

\usetikzlibrary{arrows,automata}

\usetikzlibrary{calc}

\DeclareMathAlphabet{\mathpzc}{OT1}{pzc}{m}{it}

\usepackage{epsfig}
\usepackage{graphicx}

\numberwithin{equation}{section}
\theoremstyle{plain}
\newtheorem{theorem}{Theorem}
\newtheorem{prop}[theorem]{Proposition}

\newtheorem{lemma}{Lemma}

\newtheorem{corollary}[theorem]{Corollary}
\theoremstyle{remark}
\newtheorem{remark}{Remark}[section]

\newtheorem*{quest*}{Question}
\newtheorem*{remark*}{Remark}
\theoremstyle{remark}

\theoremstyle{definition}
\newtheorem{definition}{Definition}[section]
\newtheorem*{definition*}{Definition}
\newtheorem*{notation*}{Notation}
\newtheorem*{notations*}{Notations}
\providecommand{\B}{\mathbf}

\providecommand{\D}{\mathbb}

\providecommand{\R}{\mathrm}

\newcommand{\ee}{\mathrm{e}}
\newcommand{\eu}{\mathrm{e}}

\def\ball{\mathrm{B}}

\DeclareMathOperator{\card}{card}
\DeclareMathOperator{\tr}{tr\,}

\DeclareMathOperator*{\essup}{ess\,sup}
\DeclareMathOperator*{\supp}{supp}

\DeclareMathOperator{\one}{\mathbf{1}}

\DeclareMathOperator{\Unif}{Unif}

\DeclareMathOperator{\foral}{\forall\,}

\def\mytimesL{\operatornamewithlimits{\hbox{\LARGE$\times$}}}

\def\rc{\mathrm{c}}

\def\lam{{\lambda}}

\def\prk#1{\mathbb{P}_{\Bk}\left\{ #1 \right\}}

\def\esmk#1{\D{E}_{\Bk}\left[\, #1\, \right]}

\def\rhob{\overline{\rho}}

\def\cond{\,\big|\,}

\def\BA{\mathbf{A}}

\def\BC{\mathbf{C}}

\def\BJ{\mathbf{J}}

\def\BJell{\BJ^{(\ell)}}

\def\BK{\mathbf{K}}

\def\Bk{\mathbf{k}}
\def\Bp{\mathbf{p}}

\def\Bv{\mathbf{v}}

\def\tileta{\tilde{\eta}}
\def\txi{\tilde{\xi}}
\def\teta{\tilde{\eta}}
\def\tmu{\tilde{\mu}}

\def\DP{\D{P}}
\def\DR{\D{R}}
\def\DY{\D{Y}}
\def\DZ{\D{Z}}

\def\cH{\mathcal{H}}

\def\cK{\mathcal{K}}

\def\bcK{\boldsymbol{\mathcal{\cK}}}

\def\cL{\mathcal{L}}

\def\cN{\mathcal{N}}

\def\cS{\mathcal{S}}

\def\cX{\mathcal{X}}
\def\tcX{\widetilde{\mathcal{X}}}

\def\cZ{\mathcal{Z}}

\def\cN{{\mathcal{N}}}

\def\rd{{\R{d}}}

\def\uX{\underline{X}}

\def\oX{\overline{X}}

\def\be{\begin{equation}}
\def\bel#1{\begin{equation}\label{#1}}
\def\ee{\end{equation}}

\def\ba{\begin{array}{l}}
\def\ea{\end{array}}
\def\bal{\begin{aligned}}
\def\eal{\end{aligned}}

\def\fF{\mathfrak{F}}
\def\fFbcK{\mathfrak{F}_{\boldsymbol{\mathcal{K}}}}

\def\om{{\omega}}
\def\Om{{\Omega}}

\def\Lam{{\Lambda}}
\def\lam{{\lambda}}

\def\CMnu{\B{(RCM)}}

\def\myset#1{\left\{\,#1\,\right\}}
\def\bmyset#1{\big\{\,#1\,\big\}}

\def\pr#1{\D{P}\left\{\,#1\,\right\}}
\def\esm#1{\D{E}\left[\, #1\, \right]}

\def\half{\frac{1}{2}}

\def\quart{\frac{1}{4}}

\def\myset#1{{\left\{\,#1\,\right\}}}

\def\ble{\begin{lemma}}
\def\ele{\end{lemma}}

\def\bre{\begin{remark}}
\def\ere{\end{remark}}

\def\btm{\begin{theorem}}
\def\etm{\end{theorem}}

\def\bde{\begin{definition}}
\def\ede{\end{definition}}

\def\bpr{\begin{prop}}
\def\epr{\end{prop}}

\def\bco{\begin{corollary}}
\def\eco{\end{corollary}}

\begin{document}

\title[On the conditional distribution of the sample mean]
{On the regularity of the conditional\\ distribution of the sample mean}

\author[V. Chulaevsky]{Victor Chulaevsky}


\address{D\'{e}partement de Math\'{e}matiques\\
Universit\'{e} de Reims, Moulin de la Housse, B.P. 1039\\
51687 Reims Cedex 2, France\\
E-mail: victor.tchoulaevski@univ-reims.fr}

\date{}
\begin{abstract}
We show that the hypothesis of regularity of the conditional distribution
of the empiric average of a finite sample of IID random variables,
given all the sample "fluctuations", which appeared in our earlier
manuscript \cite{C10}
in the context of the eigenvalue concentration analysis for multi-particle
random operators, is satisfied for a class of probability distributions
with sufficiently smooth probability density.
It extends the well-known property of Gaussian IID samples.
\end{abstract}

\maketitle

\section{Introduction} \label{sec:intro}

In a few talks given at workshops on disordered quantum systems, I have mentioned a simple result
of the elementary probability theory which has an interesting application to the multi-particle
Anderson localization theory. It is difficult to say if the result itself is original;
personally, I would be glad to learn that it is not, and to provide some bibliographical
reference, for it is indeed hard to believe that the elementary probabilistic problem in question
was never addressed, for example, in statistics.  However, I am unaware of any such published
(or folkloric) result.

\vskip2mm

The goal of this short note is to fill this gap and provide
an elementary proof of the
regularity (with high probability)
of the conditional sample mean of a finite sample of uniformly distributed
IID random variables, given the sigma-algebra of "fluctuations".

\vskip2mm

\textcolor{blue}{
This text is an improvement of the previous version (25.04.2013) in two ways:
\begin{itemize}
  \item we consider a larger class of probability distributions, including
  those with piecewise-constant probability density, on the intervals
  of arbitrary length $\ell$; while such a generalization is quite straightforward,
  it renders more convenient references to the main result of this paper;
  moreover, we extend the main result to a class of smooth probability densities;
\vskip2mm
  \item the probabilistic estimates are made slightly stronger; again, this is a minor
  improvement, but it may prove useful in the applications.
\end{itemize}
}

\section{Prelude: Gaussian IID samples}
\label{sec:Gauss}

Consider a sample of $N$ IID (independent and identically distributed)
random variables with Gaussian distribution $\cN(0,1)$, and
introduce the sample mean $\xi=\xi_N$ and the "fluctuations" $\eta_i$ around the mean:
$$
\xi_N = \frac{1}{N} \sum_{i=1}^N X_i,
\quad \eta_i = X_i - \xi_N, \;\; i=1, \ldots, N.
$$
It is well-known from elementary courses of the probability theory that $\xi_N$ is independent
from the sigma-algebra $\fF_\eta$ generated by $\{\eta_1, \ldots, \eta_n\}$ (the latter
are linearly dependent, and have rank $N-1$). To see this, it suffices
to note that $\eta_i$  are all orthogonal to $\xi_N$ with respect to the standard
scalar product in the linear space formed by $X_1, \ldots, X_N$ given by
$$
\langle Y, Z \rangle := \esm{ Y \, Z}
$$
where $Y$ and $Z$ are real linear combinations of $X_1, \ldots, X_N$ (recall: $\esm{X_i}=0$).

Therefore, the conditional probability distribution of $\xi_N$ given $\fF_\eta$ coincides
with the unconditional one, so $\xi_N \sim \cN(0, N^{-1})$, thus $\xi_N$ has bounded density
$$
p_\xi(t) = \frac{e^{ -\half t^2} }{ \sqrt{2\pi N^{-1}} } \le \frac{N^{1/2}}{ \sqrt{2\pi}}.
$$
Moreover, for any interval $I\subset\DR$ of length $|I|$, we have
\be\label{eq:bound.nu.Gauss}
\essup \pr{ \xi_N(\om) \in I \,\big|\, \fF} = \pr{ \xi_N(\om) \in I }
\le \frac{N^{1/2}}{ \sqrt{2\pi}} \, |I|.
\ee
The essential supremum in the above LHS is a  bureaucratic tribute to the formal
rule saying that $\pr{\,\cdot\, \,|\, \fF}$ is a random variable (which is $\fF$-measurable),
and as such is defined, generally speaking, only up to subsets of measure zero.

In this particular case -- for Gaussian samples -- the conditional regularity of the sample
mean $\xi_N$ given the fluctuations $\fF$ is granted, but is not always so, as shows the following
elementary example where the common probability distribution of the sample $X_1, X_2$
is just excellent: $X_i \sim \Unif([0,1])$, so $X_i$ admit a compactly supported
probability density bounded by $1$. Indeed, in this simple example, set
$$
\xi = \xi_2 = \frac{X_1 + X_2}{2}, \;\; \eta = \eta_1 = \frac{X_1 - X_2}{2}.
$$
The random vector $(X_1,X_2)$ is uniformly distributed in the unit square $[0,1]^2$,
and the condition $\eta = c$ selects a straight line in the two-dimensional plane
with coordinates $(X_1,X_2)$, parallel to the main diagonal $\{X_1 = X_2\}$. The conditional distribution
of $\xi$ given $\{\eta = c\}$ is the uniform distribution on the segment
$$
J_c := \{(x_1, x_2):\; x_1-x_2 = 2c,\, 0\le x_1, x_2 \le 1\}
$$
of length vanishing at $2c = \pm 1$. For $|2c|=1$, the conditional distribution of $\xi$
on $J_c$ is concentrated on a single point, which is the ultimate form of singularity.

Yet, the good news in this example is that the conditions of singularity are quite explicit,
and it is simple to assess the probability of the event that the conditional
probability density of $\xi$ given $\fF$ is bigger than a given threshold. In the next Section,
we exploit this elementary observation in a more general case of $N\ge 2$ IID
random variables uniformly distributed in $[0,1]$. The applications of the main result of Section 3
are discussed in Section 4.

\section{The principal applications}
\label{sec:applic.to.MSA}

\subsection{The conditional empirical mean in EVC bounds}
\label{ssec:role.cond.mean}

Let $\Lam$ be a finite graph, with $|\Lam|=N\ge 1$,
and $H_\Lam(\om)$ be a random DSO acting in the finite-dimensional
Hilbert space $\cH = \cH_\Lam = \ell^2(\Lam)$, with IID random potential potential $V:\Lam\times\Om\to\DR$,
relative to a probability space $(\Om,\fF,\DP)$. Decomposing the random field $V$ on $\Lam$,
$$
V(x;\om) = \xi_N(\om) + \eta_x(\om),
$$
we can represent $H(\om)$ as follows:
$$
H(\om) = \xi_N(\om) \one + A(\om),
$$
where the operator $A(\om)$ is $\fF_\eta$-measurable, and so are its eigenvalues
$\tmu_j(\om)$, $j=1, \ldots, N$. Since $A(\om)$ commutes with the scalar operator $\xi_N(\om)\one$,
the eigenvalues $\lam_j(\om)$ of $H(\om)$ have the form
\be\label{eq:lam.xi.tmu}
\lam_j(\om) = \xi_N(\om) + \mu_j(\om) .
\ee
The numeration of the eigenvalues $\lam_j(\om)$, $\mu_j(\om)$ is, of course, not canonical, but
they can be consistently defined as random variables on $\Om$.

The representation \eqref{eq:lam.xi.tmu} implies immediately the following EVC bound:
for any interval $I = [t, t+s]$,
\be
\bal
\pr{ \tr P_I(H(\om)) \ge 1} &\le \sum_{j=1}^N \pr{ \lam_j(\om) \in I}
= \sum_{j=1}^N \pr{ \xi_N(\om) + \mu_j(\om)  \in I}
\\
& = \sum_{j=1}^N \esm{ \pr{ \xi_N(\om) + \mu_j(\om)  \in I \cond \fF_\eta} }
\\
& = \sum_{j=1}^N \esm{ \pr{ \xi_N(\om) \in [- \mu_j(\om) + t, - \mu_j(\om) + t+s]\cond \fF_\eta} }
\eal
\ee
Further, omitting the argument $\om$ for notational brevity, we have
$$
\bal
\pr{ \xi_N + \tmu_j  \in I \cond \fF_\eta}
&= \pr{ \xi_N \in [ \mu_j + t,  \mu_j + t+s]\cond \fF_\eta}
\\
& = \pr{ \xi_N \in [\tmu_j, \tmu_j +s]\cond \fF_\eta}
\eal
$$
where $\tmu_j(\om) := -\mu_j(\om) + t$ are $\fF_\eta$-measurable, i.e., fixed under the
conditioning. Now introduce the conditional continuity modulus of $\xi_N$, given $\fF_\eta$:
$$
\nu_N(s) := \sup_{t\in\DR} \; \essup \; \pr{ \xi_N \in [t, t+s]\cond \fF_\eta}, \;\; s>0.
$$
Obviously,
$$
\pr{ \lam_j  \in I \cond \fF_\eta} \le \nu_N(s),
$$
thus
\be\label{eq:EVC.to.bnu}
\pr{ \tr P_I(H(\om)) \ge 1} \le N \, \nu_N(s) = |\Lam| \, \nu_N(s).
\ee

In this section, we discuss by way of example the Wegner-type bounds for a conventional,
single-particle DSO, but in applications to the multi-particle EVC bounds, similar objects
turn out to be of interest:
\be\label{eq:general.nu.mu.bound}
s \mapsto \pr{ \xi_N(\om) \in [\tmu(\om), \tmu(\om) + s  },
\ee
and
\be\label{eq:general.nu.mu.bound.cond}
s \mapsto \pr{ \xi_N(\om) \in [\tmu(\om), \tmu(\om) + s \cond \fF_\eta },
\ee
with an $\fF_\eta$-measurable random variable $\tmu$.

\subsection{The Gaussian case}
\label{ssec:cond.mean.Gauss}

In the particular case where $X_i \sim \cN(0,1)$, we can apply the estimate
\eqref{eq:bound.nu.Gauss} and infer from \eqref{eq:EVC.to.bnu} that
\be\label{eq:Wegner.nu.Gauss}
\pr{ \tr P_I(H(\om)) \ge 1} \le N \cdot \frac{N^{1/2}}{\sqrt{2\pi}}  \, |I|
= \frac{|\Lam|^{3/2}}{\sqrt{2\pi}}  \, |I|.
\ee
The above RHS gives the correct (linear) dependence upon the length of the interval $|I|$, but
the volume factor is has wrong exponent ($3/2$), compared to the Wegner estimate (with $|\Lam|^1$).
This is not surprising: we have actually exploited only one of the degrees of freedom
in the random potential, related to the normalized empirical mean $\txi_N$, while the
well-known proof, due to Wegner \cite{W81}, as well as its more recent variants, make use
of all $N=|\Lam|$ degrees of freedom. The bound \eqref{eq:Wegner.nu.Gauss} is certainly insufficient
for the proof of absolute continuity of the limiting eigenvalue distribution for the random
operator $H(\om)$ in an infinite graph (e.g., in the lattice $\DZ^d$), and this is not an intended
application of our method, as was explained in the introduction. On the other hand, it is more than sufficient for applications to the localization analysis, especially for the MSA. It would not be easy
to find an even more elementary derivation of a Wegner-like EVC bound suitable for the analysis of
resonances in disordered systems, particularly for the Gaussian potentials.

Another drawback of the described approach to the EVC estimates is that the "abstract" probabilistic
component of the proof, viz. the estimate on $\nu_N(s)$, becomes more complicated for IID ranom potentials
with low regularity of their common probability distribution function (PDF) $F_V$.
The existing methods, used in the single-particle
Anderson localization theory, provide a large choice of bounds applicable, formally, to arbitrary
continuous PDF $F_V$ (i.e., continuous marginal probability distributions); in practice, the MSA
for the DSO on lattices and more general countable graphs
requires\footnote{As it is well-known by now, owing to deep works by Bourgain--Kenig \cite{BK05},
Aizenman et al. \cite{AGKW09}, and Germinet--Klein \cite{GK13}, Anderson localization
in $\DR^d$, $d\ge 1$, can be proven for any nontrivial marginal probability distribution,
but for the discrete Schr\"{o}dinger operators this remains a challenging open problem. }
at least log-H\"{o}lder continuity of the marginal distribution. The Fractional Moments Method (FMM),
which usually provides stronger (exponential) probabilistic localization bounds, when applicable,
is even more exigent: it requires H\"{o}lder continuity of the marginal measure.

With these considerations in mind, we have to stress again that we aim here mainly at localization
analysis for multi-particle Hamiltonians, where the traditional approaches have been unable so far
to obtain efficient localization bounds in arbitrarily large finite volumes.

\subsection{Reduction to the local analysis in the sample space}

Assume that the support $\cS\subset \DR$
of the common \emph{continuous} marginal probability measure $\DP_V$ of the IID random variables
$X_j$, $1\le j \le N$, is covered by a finite or countable union of  intervals:
$$
\cS \subset \cup_{k\in \cK} J_k, \;\;\cK\subset\DZ, \; J_k = [a_k, b_k], \;\; a_{k+1} \ge b_k.
$$
Let $\BK = \cK^N$, and for each $\Bk = (k_1, \ldots, k_N)\in\BK$, denote
$$
\BJ_\Bk = \mytimesL_{i=1}^N J_{k_i}.
$$
Owing to the continuity of the marginal measure, $J_k$ are "essentially" disjoint:
for all $k\ne l$, $\DP_V(J_k \cap J_l) = 0$. Respectively, the family of the parallelepipeds
$\{\BJ_\Bk, \; \Bk\in\BK\}$ forms a partition $\bcK$ of the sample space, which we will often identify
with the probability space $\Om$. Further, let $\fFbcK$ be the sub-sigma-algebra of $\fF$
generated by the partition $\bcK$. Now the quantities of the general form
\eqref{eq:general.nu.mu.bound} can be assessed as follows:
$$
\bal
\pr{ \xi_N \in [\tmu, \tmu +s]} & = \esm{ \pr{ \xi_N \in [\tmu, \tmu +s] \cond \fFbcK} }
\\
& = \sum_{\Bk\in\BK} \pr{ \BJ_\Bk } \pr{ \xi_N \in [\tmu, \tmu +s] \cond \BJ_\Bk }.
\eal
$$
Let $\prk{\cdot}$ be the conditional probability measure, given $\{X\in\BJ_\Bk\}$,
$\esmk{\cdot}$ the respective expectation, and $p_\Bk = \pr{ \BJ_\Bk }$. Then we have
\be\label{eq:pr.xi.mu.BK}
\bal
\pr{ \xi_N \in [\tmu, \tmu +s]}
 &= \sum_{\Bk\in\BK} p_\Bk \esmk{ \prk{ \xi_N \in [\tmu, \tmu +s] \cond \fF_\eta } }
\\
& \le \sup_{\Bk\in\BK}  \esmk{ \prk{ \xi_N \in [\tmu, \tmu +s] \cond \fF_\eta } }.
\eal
\ee
This simple formula shows that one may seek a satisfactory upper bound on the LHS of
\eqref{eq:pr.xi.mu.BK} by assessing the "local" conditional probabilities
$\prk{ \xi_N \in [\tmu, \tmu +s] \cond \fF_\eta }$, where each random variable $X_j$
is restricted to a subinterval $J_{k_j}$ of its global support, so
the entire sample $X=(X_1, \ldots, X_N)$ is restricted to a parallelepiped
$\BJ \subset\DR^N$.

In the next section, we perform such analysis first in the case of a uniform marginal distribution
of the IID variables $X_i$.

\section{Uniform marginal distributions}
\label{sec:Unif}

Let be given a real number $\ell > 0$ and an integer $N\ge 2$.
Consider a sample of $N$ IID random variables with uniform distribution
$\Unif([0,\ell])$, and
introduce again the sample mean $\xi=\xi_N$ and the "fluctuations" $\eta_i$ around the mean:
$$
\xi_N = \frac{1}{N} \sum_{i=1}^N X_i,
\quad \eta_i = X_i - \xi_N.
$$
For the purposes of orthogonal transformation
$(X_1, \ldots, X_n) \mapsto (\txi_N, \teta_2, \ldots, \teta_N)$, we also need a rescaled
empirical mean
$$
\txi_N = N^{1/2} \xi_N,
$$
so
\be\label{eq:X.i.txi.eta}
X_i = \eta_i + N^{-1/2} \txi_N , \;\; i=1, \ldots N.
\ee
Further, consider the Euclidean space $\sim \DR^N$ of real linear combinations of the random variables
$X_i$ with the scalar product $\langle X', X''\rangle = \esm{X' X''}$.
Clearly, the variables $\eta_i:\DR^N\to\DR$ are invariant under the group of translations
$$
(X_1, \ldots, X_N) \mapsto (X_1 + t, \ldots, X_N+t), \;\; t\in\DR,
$$
and so are their differences $\eta_i - \eta_j \equiv X_i - X_j$, $1\le i < j \le N$.
Introduce the variables
\be\label{eq:def.Y.i}
Y_i = \eta_i - \eta_N, \; \; 1\le i \le N-1,
\ee
Then the space $\DR^N$ is fibered into a union of affine lines of the form
\be\label{eq:def.Y.i.cX}
\bal
\tcX(Y) &:= \{X\in\DR^N:\, \eta_i - \eta_N = Y_i, \, i\le N-1\}
\\
&:= \{X\in\DR^N:\, X_i - X_N = Y_i, \, i\le N-1\},
\eal
\ee
labeled by the elements $Y = (Y_1, \ldots, Y_{N-1})$ of the $(N-1)$-dimensional real
vector space $\DY^{N-1} \cong \DR^{N-1}$.
Set
$$
\cX(Y) = \tcX(Y) \cap \BC_1
= \{X\in\BC_1:\, X_i - X_N = Y_i, \, i\le N-1\}
$$
and endow each nonempty interval $\cX(Y)\subset\DR^N$
with the natural structure of a probability space inherited from $\DR^N$:
\begin{itemize}
  \item if $|\cX(Y)|=0$ (an interval reduced to a single point), then we introduce the
  trivial sigma-algebra and trivial counting measure;
  \item if $|\cX(Y)|=r>0$, then we use the inherited structure of an interval of a
  one-dimensional affine line and the normalized measure with constant density $r^{-1}$
      with respect to the inherited Lebesgue measure on $\cX(Y)$.
\end{itemize}

The transformation $X \mapsto (\xi_N, \eta_1, \ldots, \eta_{N-1})$ is non-degenerate, but
not orthogonal. We will have to work with the metric on $\cX(Y)$, induced by the standard
Riemannian metric in the ambient space $\DR^N$; to this end, introduce an orthogonal coordinate transformation in $\DR^N$
$X \mapsto (\txi_N, \tileta_1, \ldots, \tileta_{N-1})$ such that
\be\label{eq:txi.N}
\txi_N = N^{-1/2} \sum_{i=1}^N X_i = N^{1/2} \xi_N;
\ee
the exact form of $\tileta_j$, $j=1, \ldots, N-1$ is of no importance, provided
that the transformation is orthogonal.

\bre\label{rem:X.i.scaled.param}
For later use, note that, owing to \eqref{eq:txi.N}, each of the re-scaled variables $N^{1/2}X_i$ can
serve as the (normalized) length parameter on the elements $\cX(Y)$.
Along an element $\cX(Y)$, one can simultaneously parameterize $\txi$ and the variables $X_i$,
by setting $\txi(t) = c_0 + t$, $X_j(t) = c_j + N^{-1/2} t$,
with arbitrarily chosen constants $c_j$. Here, $\txi_N$ is a natural length parameter on $\cX(Y)$,
since the transformation $X \mapsto (\txi_N, \tileta_1, \ldots, \tileta_{N-1 })$ is orthogonal.
\ere

It follows from \eqref{eq:txi.N} that for any given $a\in\DR$, $s> 0$,
and some $a'\in\DR$,
\be\label{eq:xi.nu.eps}
\bal
\xi_N \in [a, a + s] &\Longleftrightarrow
 \txi_N \in[a', a' + N^{1/2}s]
\eal
\ee

Next, denote $\BJ^{(\ell)} = [0,\ell]^N$ and introduce the random variable
\be\label{eq:nu.s}
\bal
\nu_N(s; \BJell) = \nu_N(s;\BJell; X) &:=
\essup\; \sup_{t\in\DR} \pr{ \xi_N\in[t, t + s] \,\big|\, \fF_\eta} .
\eal
\ee
Here the presence of $\essup$ is the tribute to the fact that the conditional probabilities are random
variables, usually defined up to subsets of zero measure; $\ell>0$ is the width of the common uniform
distribution of $X_j$. Equivalently, one may write $\nu_N(s;\BJell; \om)$ instead of
$\nu_N(s;\BJell; X)$, since the sample space $\DR^N$ is identified with the underlying
probability space $\Om$.

Since $\{X_i\}$ are IID with uniform distribution on $[0,\ell]$, the distribution of the random
vector $X(\om)$ is uniform in the cube $\BJell=[0,\ell]^N$, inducing a uniform conditional distribution
on each element $\cX(Y)$. Therefore, by \eqref{eq:xi.nu.eps} and \eqref{eq:nu.s},
\be\label{eq:nu.eps.1}
\bal
\nu_N(s;\BJell) = \frac{ N^{1/2} s}{ |\cX(Y) | } .
\eal
\ee

It is to be stressed that both sides of the above equality are random variables:
$\nu_N(s;\ell) = \nu_N(s;\ell;\om)$ by its definition in \eqref{eq:nu.s}, and
$\cX(Y) = \cX(Y(X(\om)))$.

\ble\label{lem:nu.xi.unif.ell}
Consider the IID random variables $X_1, \ldots, X_N$ with $X_i \sim \Unif(J_{\ell,i})$,
where $J_{\ell,i} = [a_i, a_i + \ell] \subset\DR$ , $\ell>0$.
For any $0 < \delta \le \ell$,
\be\label{eq:lem.nu.xi.unif.ell.1}
\bal
\pr{ |\cX(X)| \le \delta } &\le
\sum_{i=1}^N \pr{X_i -a_i < \delta } .
\eal
\ee
\ele

\proof
Without loss of generality, we can consider the case where $a_i = 0$, $1\le i \le N$,
so $X_i \sim \Unif([0,\ell])$. Otherwise, we make change of variables $X_i \mapsto X_i - a_i$.

Let
\be\label{eq:def.X.star}
\uX = \uX(X) = \min_{i} X_i.
\ee
According to Remark \ref{rem:X.i.scaled.param}, each $N^{1/2}X_i$, $i=1, \ldots, N$,
restricted to $\cX(Y)$, provides a normalized length parameter on $\cX(Y)$; thus the range
of each $N^{1/2} X_i |_{\cX(Y)}$ is an interval of length $|\cX(Y)|$. One can decrease,
e.g., the value of $X_1$, as long as \emph{all} $\{X_i, 1\le i \le N\}$ are strictly strictly positive. Therefore, the maximum decrement of $X_1$
(indeed, of any $X_i$) along $\cX(Y)$ is given by $\uX(X)$, so the range
of the normalized length parameter $N^{1/2} X_1$ along $\cX(Y(X))$ is
an interval of length $ \ge N^{1/2}\uX(X)$:
\be\label{eq:len.cX.oX.uX}
|\cX(Y(X))| \ge N^{1/2} \uX(X).
\ee
Let
\be\label{eq:A.i.i.empty.1}
 A_{i}(t) := \{  X_i < t \}, \;\; \BA(t) := \cup_{i=1}^N \; A_i(t), \;
 \BA^\rc(t) = \Om \setminus\BA(t),
\ee
and note that, by \eqref{eq:len.cX.oX.uX},
$$
\min_{ X \in \BA^\rc(t)} |\cX(X)|
\ge N^{1/2} \min_{ X \in \BA^\rc(t)} \uX(X) \ge N^{1/2} t.
$$
Equivalently, setting $u = N^{1/2} t$, so $t = N^{-1/2}u$, we have
\be
|\cX(X)| < u \;\; \Longrightarrow \;\; X \in \BA(N^{-1/2} u).
\ee
With $u = \delta$, we infer from
\eqref{eq:nu.eps.delta.1.lem}
\be\label{eq:prob.Ai.to.r}
\bal
\pr{ \BA\left( N^{1/2} N^{-1/2}  \delta \right) }
&= \pr{ \BA\left( \delta \right) }
\le \sum_{i=1}^N \pr{ X_i < \delta }  .
\eal
\ee
proving the assertion \eqref{eq:lem.nu.xi.unif.ell.1}.
\qedhere

\vskip3mm

\btm\label{thm:nu.xi.unif.ell.linear}
Consider IID random variables $X_1, \ldots, X_N$ with $X_i \sim \Unif(J_{\ell,i})$,
where $J_{\ell,i} = [a_i, a_i + \ell] \subset\DR$ , $\ell>0$.
For any $0 < \delta\le \ell$,
\be\label{eq:thm.nu.xi.unif.ell.lin.1}
\bal
\pr{ \nu_N(s;\BJell) > \delta^{-1} s } &\le \frac{ N \delta}{\ell} .
\eal
\ee
In particular, with $\delta = s^{\alpha}$,
\be\label{eq:thm.nu.xi.unif.ell.lin.2}
\pr{ \nu_N(s;\BJell) >  s^{1-\alpha} } <  N \ell^{-1} s^{\alpha}
\ee
\etm
\proof

The random variable
$
X=(X_1, \ldots, X_N) \mapsto |\cX(Y(X))|
$
is $\fF_\eta$-measurable and takes constant value $|\cX(Y)|$
on each element $\cX(Y)$.
By \eqref{eq:nu.eps.1}, for any $\delta>0$,
\be\label{eq:nu.eps.delta.1.lem}
\pr{ \nu_N(s; \BJell) \ge \delta^{-1} s } \le \pr{ \frac{N^{1/2} s}{|\cX(Y)|} \ge \delta^{-1} s }
= \pr{ |\cX(Y)| \le N^{1/2} \delta }.
\ee
Now \eqref{eq:thm.nu.xi.unif.ell.lin.1} follows from \eqref{eq:nu.eps.delta.1.lem} and
Lemma \ref{lem:nu.xi.unif.ell}, since for $X_i \sim \Unif([0,\ell])$
$$
\pr{X_i < \delta} = \ell^{-1} \delta.
$$
\qedhere

\section{More accurate bounds}

A direct inspection shows that the bounds of Lemma \ref{lem:nu.xi.unif.ell}
(and, consequently, those of Theorem \ref{thm:nu.xi.unif.ell.linear})
are not optimal, since they are based on the inequality
\be\label{eq:len.cX.oX.uX.again}
|\cX(Y(X))| \ge N^{1/2} \uX(X)
\ee
(cf. \eqref{eq:len.cX.oX.uX.again}) which can be easily improved; we do so in Theorem
\ref{thm:nu.xi.unif.ell} below. However, the method of proof of Lemma \ref{lem:nu.xi.unif.ell}
is simpler and quite sufficient for our main application to the multi-particle MSA.

\ble\label{lem:prob.small.cX.density}
Assume that the IID random variables $X_1, \ldots, X_N$, $N\ge 2$, admit (common) probability density
$p_V$ with $\|p_V\|_\infty \le \rhob<\infty$. Then
\be\label{eq:lem.small.cX.density.1}
\pr{ |\cX(Y)| < r } \le \frac{1}{4} \rhob^2 r^2 N.
\ee
In particular, for $X_j \sim \Unif([0,\ell))$, one has
\be\label{eq:lem.small.cX.density.2}
\pr{ |\cX(Y)| < r } \le \frac{r^2 N}{4 \ell^2}.
\ee
\ele
\proof

Let
\be\label{eq:def.X.star}
\uX = \uX(X) = \min_{i} X_i, \; \oX = \oX(X)=\max_{i} X_i.
\ee
While $\oX(X)$ and $\uX(X)$ vary along the elements $\cX(Y)$, their difference
$\oX(X) -\uX(X)$ does not; it is uniquely determined by $\cX(Y)$.

According to Remark \ref{rem:X.i.scaled.param}, each $N^{1/2}X_i$, $i=1, \ldots, N$,
restricted to $\cX(Y)$, provides a normalized length parameter on $\cX(Y)$; thus the range
of each $N^{1/2} X_i |_{\cX(Y)}$ is an interval of length $|\cX(Y)|$. One can increase (resp., decrease),
e.g., the value of $X_1$, as long as \emph{all} $\{X_i, 1\le i \le N\}$ are strictly smaller than $\ell$
(resp., strictly positive). Therefore, the maximum increment of $X_1$
(indeed, of any $X_i$) along $\cX(Y)$
is given by $\ell - \oX(X)$, and its maximum decrement equals $\uX(X)$, so the range
of the normalized length parameter $N^{1/2} X_1$ along $\cX(Y(X))$ is
an interval of length $N^{1/2}\big(\ell - \oX(X) + \uX(X) \big)$:
\be\label{eq:len.cX.oX.uX.again}
|\cX(Y(X))| = N^{1/2} \big( \ell - \oX(X) + \uX(X) \big),
\ee
Since both $\uX(X)$ and $\ell -\oX(X)$ are non-negative,
\be\label{eq:def.r.Y}
 \uX + (\ell -\oX) < t  \;\; \Longrightarrow \;\; \max\{\uX,\; \ell - \oX\} < t/2.
\ee
With $0 \le t\le \ell$, $\big(\ell-X_i < t/2\big)$ implies $\big(X_i > t/2\big)$, thus
denoting
\be\label{eq:A.i.i.empty.1}
 A_{ij}(t) := \{  X_i < t/2 \} \cap \{ \ell-X_j < t/2 \},
\ee
we have, for any $i$,
\be\label{eq:A.i.i.empty.2}
A_{ii}(t) = \{ X_i < t/2 \}  \, \cap \, \{ \ell-X_i < t/2 \}
= \varnothing.
\ee
Therefore,
\be\label{eq:sum.Akij.unif}
\left\{\max\big\{\uX(X),\; \ell-\oX(X)\big\} < \frac{t}{2}  \right\} \subset
\bigcup_{i \ne j} \left\{  X_i < \frac{t}{2}, \; \ell-X_j < \frac{t}{2} \right\}.
\ee
Thus the union $\cup_{i\ne j} A_{ij}(t)$ contains all samples $X$ with $|\cX(Y)| < t/2$.

The sample $\{X_k\}$ is IID, with common probability density uniformly bounded by $\rhob<\infty$,
so for any $i\ne j$
$$
\pr{A_{ij}(t)} = \pr{ X_i < \frac{t}{2}} \cdot \pr{ \ell-X_j < \frac{t}{2} }
= \frac{1}{4} \rhob^2 t^2.
$$
Therefore,
\be\label{eq:prob.Aij.to.r}
\bal
\pr{ |\cX(Y)| < r} &= \pr{ N^{1/2} \big( (\ell - \oX(X)) + \uX(X) \big) < r }
\\
& = \pr{ \big( (\ell - \oX(X)) + \uX(X) \big) < r N^{-1/2} }
\\
&
\le \sum_{i \ne j} \pr{ A_{ij}\big( r N^{-1/2} \big) }
  \le  N(N-1) \, \frac{ \left(\rhob  r N^{-1/2}\right)^2}{ 4 }
\\
& \le \frac{1}{4} \rhob^2 r^2 N.
\eal
\ee

\qedhere

\btm\label{thm:nu.xi.unif.ell}
Consider the IID random variables $X_1, \ldots, X_N$ with $X_i \sim \Unif([0,\ell])$.
For any $0 < \delta \le s \le \ell$,
\be\label{eq:thm.nu.xi.unif.ell.1}
\bal
\pr{ \nu_N(s;\BJell) > \delta^{-1} s } &\le \frac{N^2 \delta^2}{4\ell^2} .
\eal
\ee
In particular, with $\delta =  s^{\alpha}$, $\alpha\in(0,1)$,
\be\label{eq:thm.nu.xi.unif.ell.2}
\pr{ \nu_N(s;\BJell) >  s^{1-\alpha} } <  \frac{N^2s^{2\alpha}}{ 4\ell^2}
\ee
\etm

\proof
As before, we associate with each point $X\in\DR^N$ the straight line $\cL(Y(X))\ni X$
parallel to the vector $\Bv = (1, \ldots, 1)$.
and consider their intersections $\cX(Y(X)) = \cL(Y(X)) \cap \BJell$.
Owing to Eqn \eqref{eq:nu.s}, for any $\delta>0$,
\be\label{eq:nu.eps.delta.1}
\pr{ \nu_N(s) \ge \delta } \le \pr{ \frac{N^{1/2} s}{|\cX(Y)|} \ge \delta }
= \pr{ |\cX(Y)| \le N^{1/2} s \delta^{-1}}.
\ee
Let
\be\label{eq:def.X.star}
\uX = \uX(X) = \min_{i} X_i, \; \oX = \oX(X)=\max_{i} X_i.
\ee
While $\oX(X)$ and $\uX(X)$ vary along the elements $\cX(Y)$, their difference
$\oX(X) -\uX(X)$ does not; it is uniquely determined by $\cX(Y)$.

According to Remark \ref{rem:X.i.scaled.param}, each $N^{1/2}X_i$, $i=1, \ldots, N$,
restricted to $\cX(Y)$, provides a normalized length parameter on $\cX(Y)$; thus the range
of each $N^{1/2} X_i |_{\cX(Y)}$ is an interval of length $|\cX(Y)|$. One can increase (resp., decrease),
e.g., the value of $X_1$, as long as \emph{all} $\{X_i, 1\le i \le N\}$ are strictly smaller than $\ell$
(resp., strictly positive). Therefore, the maximum increment of $X_1$
(indeed, of any $X_i$) along $\cX(Y)$
is given by $\ell - \oX(X)$, and its maximum decrement equals $\uX(X)$, so the range
of the normalized length parameter $N^{1/2} X_1$ along $\cX(Y(X))$ is
an interval of length $N^{1/2}\big(\ell - \oX(X) + \uX(X) \big)$:
\be\label{eq:len.cX.oX.uX.again}
|\cX(Y(X))| = N^{1/2} \big( \ell - \oX(X) + \uX(X) \big),
\ee
Since both $\uX(X)$ and $\ell -\oX(X)$ are non-negative,
\be\label{eq:def.r.Y}
 \uX + (\ell -\oX) < t  \;\; \Longrightarrow \;\; \max\{\uX,\; \ell - \oX\} < t/2.
\ee
With $0 \le t\le \ell$, $\big(\ell-X_i < t/2\big)$ implies $\big(X_i > t/2\big)$, thus
denoting
\be\label{eq:A.i.i.empty.1}
 A_{ij}(t) := \{  X_i < t/2 \} \cap \{ \ell-X_j < t/2 \},
\ee
we have, for any $i$,
\be\label{eq:A.i.i.empty.2}
A_{ii}(t) = \{ X_i < t/2 \}  \, \cap \, \{ \ell-X_i < t/2 \}
= \varnothing.
\ee
Therefore,
\be\label{eq:sum.Akij.unif}
\left\{\max\big\{\uX(X),\; \ell-\oX(X)\big\} < \frac{t}{2}  \right\} \subset
\bigcup_{i \ne j} \left\{  X_i < \frac{t}{2}, \; \ell-X_j < \frac{t}{2} \right\}.
\ee
Thus the union $\cup_{i\ne j} A_{ij}(t)$ contains all samples $X$ with $|\cX(Y)| < t/2$.

The sample $\{X_k\}$ is IID, with $X_k \sim \Unif([0,\ell])$,
so for any $i\ne j$
$$
\pr{A_{ij}(t)} = \pr{ X_i < \frac{t}{2}} \cdot \pr{ \ell-X_j < \frac{t}{2} }
= \frac{t^2}{4\ell^2}
$$
Owing to \eqref{eq:len.cX.oX.uX},
\be\label{eq:prob.Aij.to.r}
\bal
\pr{ |\cX(Y)| < r} &= \pr{ N^{1/2} \big( (\ell - \oX(X)) + \uX(X) \big) < r }
\\
& = \pr{ \big( (\ell - \oX(X)) + \uX(X) \big) < r N^{-1/2} }
\\
&
\le \sum_{i \ne j} \pr{ A_{ij}\big(r N^{-1/2} \big) }
  \le  N(N-1) \, \frac{ \left(r N^{-1/2}\right)^2}{ 4 \ell^2}
\\
& \le \frac{ r^2 N}{4\ell^2},
\eal
\ee
Setting $r = N^{1/2}  \delta$, we infer from \eqref{eq:nu.eps.delta.1}
\be\label{eq:prob.Aij.to.r}
\pr{ \nu(s;\ell) > \delta } \le \frac{ N^2 s^2 }{ 4 \ell^2} .
\ee
proving \eqref{eq:thm.nu.xi.unif.ell.1}. The estimate
\eqref{eq:thm.nu.xi.unif.ell.2} is a particular case
of \eqref{eq:thm.nu.xi.unif.ell.1}.
\qedhere


\vskip4mm

In Ref. \cite{C10}, we introduced the following more general condition, which actually does not
assume the independence of the random variables $X_j$. Let us reformulate it now in
a more general way so as to adapt it to locally finite connected graphs $\cZ$ with polynomially bounded
growth of balls $\ball_L(u) := \{x\in\cZ:\; \rd_\cZ(x,u)\le L\}$  (in \cite{C10}, we had $\cZ = \DZ^d$):
\be\label{eq:cond.growth.balls}
\card\, \ball_L(u) \le C_d L^d, \; l\ge 1.
\ee
(We also adapt the notation of  \cite{C10} to match the one used in this paper.)

Let $Q\subset \ball_R(x)\subset\cZ$ be a subset of a ball of radius $R$.
Consider the sample of IID random variables
$\{ V(y;\om), \, y\in Q \}$; introduce as in \eqref{eq:nu.s}
the sample mean
$\xi_{Q}$ and the conditional continuity modulus
$\nu_{|Q|}(s)$ given the sigma-algebra of fluctuations.
Since $Q\subset \ball_R(x) \subset \cZ$, where $\cZ$ satisfies
\eqref{eq:cond.growth.balls}, we have $|Q|\le C_d R^d$.

The hypothesis used in \cite{C10}, reformulated for general index sets $Q$,
takes the following form: for some
$C', C'', A', A'', B', B''\in(0,+\infty)$
\be\label{eq:CMnu.recall}
\pr{ \nu_{|Q|}(s) \ge C' |Q|^{A'} s^{B'} }  \le C'' |Q|^{A''} s^{B''}.
\ee
To  keep track of the length $\ell$ of the interval $[0,\ell]$, re-write it
as follows:
\be\label{eq:CMnu.recall.ell}
\pr{ \nu_{|Q|}(s;\ell) \ge C'|Q|^{A'} s^{B'} } \le C'' |Q|^{A''} s^{B''}.
\ee

We will say that a random field $V:\cZ\times\Omega\to\DR$ on a countable set $\cZ$
(not necessarily IID) is of class $\CMnu$ (here $RCM$ stands for
"\emph{Regularity of the Conditional Mean}")
if it satisfies the condition \eqref{eq:CMnu.recall.ell} for some values
$C', C'', A', A'', B', B''\in(0,+\infty)$. Naturally, it can be made less cumbersome,
since some of these constants can be eliminated by a proper scaling of the variable $s$,
but it might be convenient in some applications to keep all these parameters.

If the random field $V$ is assumed IID, then \eqref{eq:CMnu.recall.ell} is merely
a condition on the common marginal probability distribution; in this particular (but important) case,
one can speak of the class $\CMnu$ of the probability distributions.

We see that, for an IID sample with  distribution $\Unif([0, \ell])$, $\ell>0$,
Theorem \ref{thm:nu.xi.unif.ell} can be reformulated in the following way:

\btm\label{thm:unif.ell.implies.CMnu}
Let an IID random field $V:\cZ\times\Om$ on a finite or countable graph
$\cZ$, satisfying the growth condition \eqref{eq:cond.growth.balls},
have marginal distribution $\Unif([c, c+\ell])$, $c\in\DR$.
Then $V$ satisfies
the condition $\CMnu$ of the form \eqref{eq:CMnu.recall.ell} with the parameters
which can be chosen as follows:
\be
\bal
C' &= 1,  &\;A' &= 0,  \;&b' &= 1 -\alpha, \;\;
\\
C'' &= \frac{1}{4\ell^2}, \;&A'' &= 2, &\;b'' &= 2\alpha.
\eal
\ee
For example, one can set
\be
\bal
b'  = b'' = 2/3.
\eal
\ee
Explicitly,
\be\label{eq:thm.nu.xi.unif.ell.2.again}
\pr{ \nu_{|Q|}(s;\ell) >  s^{1-\alpha} } < \frac{|Q|^2}{4 \ell^2} s^{2\alpha} .
\ee

\etm


\section{Smooth positive densities }
\label{sec:smooth}

Now we consider a richer class of probability distributions. While the conditions
which we will assume are certainly very restrictive (uniform positivity
and smoothness of the \emph{probability density} on a compact interval), they are quite sufficient
for applications to physically realistic Anderson models.

A direct inspection of the proof of Theorem \ref{thm:nu.xi.ell.piecewise}
evidences that the hypothesis of strict positivity of the probability density
($\rho \ge \rho_*>0$, cf. \eqref{eq:cond.rho.ge.s.b} below) can be easily replaced
by a more general condition of mild decay at the endpoints of $\supp\, \rho$, e.g.,
$$
\rho(t) \ge C \left( \min\{t, \ell-t\} \right)^a, \;\; C,a\in(0,+\infty).
$$
This extends our result to a large class of popular a.c. probability distributions,
including the convolution powers of the uniform distribution. Further, the distributions with
unbounded support can be treated as well, provided that the probability density decays sufficiently fast 
at infinity (e.g., the exponential distribution and, more generally, gamma-distributions).
We plan to address such probability measures in a forthcoming paper.

\btm\label{thm:nu.xi.ell.piecewise}
Assume that the common probability distribution of the IID random variables
$V_j, \, j=1, \ldots, N$,
with  PDF $F_V$,  satisfies the following conditions:
\begin{enumerate}[label=\rm(\roman*),leftmargin=2.1em,align=right]
  \item the probability distribution is absolutely continuous:
\be\label{eq:cond.rho.ge.s.a}
dF_V(v) = \rho(v)\, dv, \; \supp \rho = [0,\ell];
\ee
  \item there exist $\rho_*, \rhob\in(0,+\infty)$ such that
\be\label{eq:cond.rho.ge.s.b}
 \forall\, t\in [0, \ell] \quad
\rho_* \le  \rho(t) \le \rhob ;
\ee
  \item
$\rho$ has bounded derivative on $(0, \ell)$:
\be\label{eq:cond.rho.ge.s.c}
 \left\| {\rho'}(\cdot) \, {\one_{(0, \ell)}}\right\|_\infty \le C'_{\rho} < +\infty.
\ee
\end{enumerate}
Then there exists $c_* = c_*(F_V)>0$ such that for any
$\delta \in \left(0, c_* N^{-3/2} \right]$,
\be\label{eq:thm.nu.xi.unif.ell.smooth.1}
\bal
\pr{ \nu_N(s ) > \delta^{-1} s } & < \frac{ 4\rhob^2 N^2\, \delta^2}{ \ell^2} .
\eal
\ee
In particular, with $\delta = s^{\alpha} \le c_*^{1/\alpha} N^{-3/(2\alpha)}$, $\alpha\in(0,1)$, one has
\be\label{eq:thm.nu.xi.unif.ell.smooth.2}
\bal
\pr{ \nu_N(s ) > s^{1-\alpha} } & < \frac{ 4\rhob^2 }{  \ell^2} N^2 s^{2\alpha}.
\eal
\ee
Consequently, the IID random fields satisfying
(i)--(iii)
belong to the class  $\CMnu$.
\etm

\proof

\par\vskip2mm\noindent
\textbf{Step 1. Smoothness of the conditional measure.}
Unlike the model considered in Section \ref{sec:Unif},
the conditional probability distribution induced on a given interval $\cX(Y)$ is
no longer constant. However, owing to the smoothness assumption (iii),
the product probability measure with density
$$
\Bp(x_1, \ldots, x_n) = \prod_{j=1}^n \rho(x_j) = \eu^{ \sum_{j=1}^n \ln \rho(x_j) }
$$
induces on the interval $\cX(Y)\subset\cL(Y)$ a measure with smooth density with respect to the  Lebesgue measure on the line $\cL(Y)\subset\DR^N$. Let $t = \txi_N$ be the
normalized length parameter along $\cL(Y)$, then (cf. \eqref{eq:X.i.txi.eta})
$$
\cL(Y) = \big\{ \big(\eta_1 + t N^{-1/2},  \ldots, \eta_N + tN^{-1/2} \big), \; t\in \DR  \big\} ,
$$
so the density at the point $t$ has the form
$$
p(t)
= Z^{-1}(Y) \prod_{j=1}^n \rho( \eta_j + t) = \eu^{ \sum_{j=1}^n \ln \rho( \eta_j + t ) }
$$
where $Z^{-1}(Y)$ is the normalization factor.
In particular,
\be\label{eq:log.deriv.p}
\frac{d}{d t} p(t)
= N^{-1/2} \, p(t) \sum_{j=1}^N \frac{ \rho'( \eta_j + t N^{-1/2} ) } { \rho( \eta_j + t N^{-1/2} ) } .
\ee

\par\vskip2mm\noindent
\textbf{Step 2. From $\nu$ to $|\cX(Y)|$.}
By \eqref{eq:log.deriv.p} combined with assumption \eqref{eq:cond.rho.ge.s.b},
$$
\left\| \frac{p'}{ p } \,\Big|_{\cX(Y)} \right  \|_\infty  \le
N \cdot N^{-1/2} C'_{\rho} \rho_*^{-1}  \le C_1 N^{1/2},
$$
In particular,
\be\label{eq:deriv.norm}
\| p' |_{\cX(Y)} \| \le C_1 N^{1/2} \, \big\| p |_{Y} \big\|_\infty.
\ee
For notational convenience, identify $\cL(Y)$ with the real line $\DR$,
equipped with the normalized coordinate $t=\txi_N$,
and let $t^* = t^*(Y)$
be any point of maximum of the density $\rho$ restricted to $\cX(Y)$, and
$\rho^*(Y) = \rho(t^*)$;
the existence of $t^*(Y)$ follows from the continuity of $\rho$.
Assume that
\be\label{eq:def.ell.N}
 |\cX(Y)| >  2\ell_N, \;\; \ell_N \le \ell_* N^{-1},
\ee
where $\ell_* = \ell_*(F_V)>0$ is small enough, viz.
$$
 \ell_*(F_V) = (C_1(F_V))^{-1},
$$
and depends upon the minimum of the density $p(\cdot)$ and the sup-norm of its derivative;
both of these quantities are determined by the PDF $F_V$.
Since $|\cX(Y)| >  2\ell_N$, at least one of the intervals $[t^* - \ell_N, t^*$,
$[t^*, t^* + \ell_N$ (perhaps, both of them) is inside the interval $\cX(Y)$.
denote by $J_{*}$ such an intervals (for definiteness, the first one, if both are inside $\cX(Y)$).

Then for any $t\in \cX(Y)$, owing to \eqref{eq:deriv.norm},
$$
\bal
\big| \rho(t) - \rho(t^*) \big| &\le  \ell_* N^{-1} \cdot \max_{s\in J_{*}} \rho'(s)
\le \underbrace{\big( C_1 \ell_* \big)}_{\le 1} N^{1/2} \cdot N^{-1} \cdot \rho^*(Y)
\eal
$$
so that $\foral t\in\cX(Y)$ and, e.g., $N\ge 4$,
$$
\half \rho^*(Y) \le \rho^*(Y) \big(1 - N^{-1/2} \big) \le  \rho(t) \le \rho^*(Y) \big(1 + N^{-1/2} \big)
\le 2 \rho^*(Y)
$$
The conditional mesure induced on $\cX(Y)$ has the form $dP_Y(t) = Z^{-1}(Y) \rho(t) \, dt$, with
$Z(Y) = \int_{\cX(Y)} \rho(t)\, dt$, and we have
$$
Z(Y) \ge \int_{J_*} \half \rho(t^*) \, dt = \half \rho(t^*) \ell_N.
$$
Therefore, under the assumption $|\cX(Y)| \ge 2\ell_N$, we have for any $t'\in\DR$:
$$
\bal
\pr{ \xi_N \in [t', t'+s] \cond Y } &=
\pr{ \txi_N \in [t'', t''+  N^{1/2} s] \cond Y }
\\
&= Z^{-1}(Y) \int_{t''}^{t''+ N^{1/2}s} \rho(t) \, dt
\\
& \le \frac{ \rho(t^*) N^{1/2} s }{ \rho(t^*) \ell_N/2}
=  \frac{2 N^{1/2} }{ \ell_N} \, s
\eal
$$
(here $t'' = N^{1/2} t'$),
yielding, for such $\cX(Y)$,
$$
\bal
\nu_N(s \cond Y) &\le 2 N^{1/2} \ell_N^{-1} s.
\eal
$$
Therefore,
$$
\myset{ \nu_N(s) > 2 N^{1/2} \ell_N^{-1} s } \subset \bmyset{ |\cX(Y)| < 2 \ell_N }
$$
Set $\delta := \half N^{-1/2} \ell_N$, $c_*=c_*(F_V) := \half \ell_*(F_V)$. Then
for any $0 <\delta \le c_* N^{-3/2}$,
\be\label{eq:nu.cX.delta}
\myset{ \nu_N(s) > s \delta^{-1} } \subset \bmyset{ |\cX(Y)| < 4 N^{1/2} \delta }.
\ee

\par\vskip2mm\noindent
\textbf{Step 3. Conclusion.}
Now we apply Lemma \ref{lem:prob.small.cX.density}
(cf. \eqref{eq:lem.small.cX.density.1}),
$$
\bal
\pr{ |\cX(Y)| <  r  } \le \quart \rhob^2 r^2 N,
\eal
$$
and obtain, with $r = 4 N^{1/2} \delta$,
\be\label{eq:prob.cX.delta}
\pr{ |\cX(Y)| < 4 N^{1/2} \delta } \le 4  \rhob^2 N^2\, \delta^2.
\ee
Now the main assertion follows from \eqref{eq:prob.cX.delta} and \eqref{eq:nu.cX.delta}:
for any $\delta \in \left(0, c_* N^{-3/2} \right)$
$$
\bal
\pr{ \nu_N(s) > \delta^{-1} s } \le 4  \rhob^2 N^2\, \delta^2.
\eal
$$
\qedhere


\end{document}